\documentclass{article}

\usepackage{amssymb,amsfonts,amsmath}
\usepackage{cite,enumerate,float,indentfirst}
\usepackage{color}

\def\be{\begin{eqnarray}}
\def\ee{\end{eqnarray}}
\def\nn{\nonumber}

\def\p{\partial}
\def\tr{{\rm tr}\,}
\def\Tr{{\rm Tr}\,}

\def\Mac{M}
\def\Mac{{\rm Mac}}
\def\Schur{{\rm Schur}}
\def\SchurP{{\cal P}}
\def\SchurP{{\rm Ker}}
\def\Ker{\SchurP}

\definecolor{red}{rgb}{1,0,0}
\definecolor{orange}{rgb}{1,0.5,0}
\definecolor{violet}{rgb}{0.7,0,1}

\hoffset= - 1.0in         

\begin{document}

\title{\vspace{-.5cm}{\Large {\bf  Kerov functions revisited}\vspace{.2cm}}
\author{
{\bf A.Mironov$^{a,b,c}$}\footnote{mironov@lpi.ru; mironov@itep.ru}\ \ and
\ {\bf A.Morozov$^{b,c}$}\thanks{morozov@itep.ru}}
\date{ }
}

\maketitle

\vspace{-5cm}

\begin{center}
\hfill FIAN/TD-21/18\\
\hfill IITP/TH-19/18\\
\hfill ITEP/TH-33/18
\end{center}

\vspace{2.3cm}

\begin{center}
$^a$ {\small {\it Lebedev Physics Institute, Moscow 119991, Russia}}\\
$^b$ {\small {\it ITEP, Moscow 117218, Russia}}\\
$^c$ {\small {\it Institute for Information Transmission Problems, Moscow 127994, Russia}}

\end{center}

\vspace{.5cm}

\begin{abstract}
The Schur functions play a crucial role in the modern description of
HOMFLY polynomials for knots and of topological vertices in DIM-based
network theories, which could
merge into a unified theory still to be developed.
The Macdonald functions do the same for hyperpolynomials and
refined vertices, but merging appears to be more problematic.
For a detailed study of this problem, more knowledge is needed
about the Macdonald polynomials than is usually available.
As a preparation for the discussion of the knot/vertices relation,
we summarize the relevant facts and open problems about the Macdonald
and, more generally, Kerov functions.
Like Macdonald polynomials, they are triangular combinations of Schur
functions, but orthogonal in a more general scalar product.
We explain that parameters of the measure can be considered as a set of
new time variables, and the Kerov functions are actually expressed
through the Schur functions of these variables as well.
Despite they provide an infinite-parametric extension of the Schur and Macdonald
polynomials, the Kerov functions, and even the skew Kerov functions
continue to satisfy the most important relations, like Cauchy
summation formula, the transposition identity for reflection of the Young diagram
and expression of the skew functions through the generalized Littlewood-Richardson
structure constants.
Since {\it these} are the properties important in most applications,
one can expect that the Kerov extension exists for most of them,
from the superintegrable matrix and tensor models to knot theory.
\end{abstract}

\vspace{.0cm}

\section{Introduction}

Recently in \cite{L8n8} a program was originated
to construct various (not obligatory torus) link polynomials \cite{knotpols}
from topological vertices  \cite{topvert}.
Already in the simplest example of the link $L_{8n8}$
considered in that paper,
one encounters composite representations,
associated "uniform" Schur functions and their
decompositions into bilinear combinations of the
more conventional skew Schur functions,
and the skew Schur is what is actually associated with the topological vertex.

This technique is however not easily lifted from the HOMFLY polynomials to
the level of hyperpolynomials
(the algebraically defined counterparts
of Khovanov-Rozansky and superpolynomials),
largely because of our insufficient knowledge about the
Macdonald polynomials, which substitute the Schur functions
in this lifting.
This paper is a preparation for developing a theory of composite
Macdonald functions and its application to study
of the hyperpolynomial/refined-vertex relation,
which will be presented elsewhere \cite{NagoyaITEP}.
Needed for  this theory are the basic facts about
the Macdonald functions in the form which allows one to easily
calculate them for arbitrary representations $R$ at sufficiently
high levels $|R|$.

In this paper, we make such a list of properties,
moreover, we present it for more general Kerov functions \cite{Kerov},
because in applications it is important to understand what in
these functions is depending on
the obstacles for the
Schur $\longrightarrow$ Macdonald $\longrightarrow$ Kerov
extension, what are the ways to overcome these obstacles
and what one needs to sacrifice when doing this.

We begin with the most general Kerov functions and then
descend to the Macdonald and Schur functions, which have some
additional properties absent in the Kerov case.
The very few properties, which are true only for the Schur functions
and are destroyed already by the Macdonald deformation are underlined,
those which are true for the Macdonald, but not for generic Kerov functions
are underlined twice.
The Macdonald functions have closer relation to group theory
but many of their properties
are actually true in the more general Kerov framework.

The Kerov {\it function} should not be confused
with much better known  "Kerov character polynomials",
also associated with the name of S.Kerov \cite{Kerpols}:
to avoid the confusion, we use the term {\it functions}
throughout the present text.

\section{  Kerov functions and their properties}

\begin{itemize}

\item{
The pair of Kerov functions $\SchurP_R^{(g)}\{p\}$ and $\widehat{\SchurP}_R^{(g)}\{p\}$
depends on  the Young diagram (representation) $R=[r_1\geq r_2\geq \ldots \geq r_{l_R}>0]$,
they are polynomials of infinitely many time-variables $p_k$ homogeneous w.r.t. the grading ${\rm grad}(p_k)=k$. Thus, a
particular $\SchurP_R\{p\}$ depends only on $|R|$ times,
{\it the level} $|R|$ being the size of $R$ (number of boxes).
The two functions $\SchurP_R^{(g)}\{p\}$, $\widehat{\SchurP}_R^{(g)}\{p\}$ differ by the choice of ordering in the sums in (\ref{PthroughSchurs}) below.
}

\item{
The best for practical purposes is to define the Kerov and Macdonald functions
by a {\it triangular} transform from the Schur functions,
\be
\boxed{
\left\{
\begin{array}{c}
\SchurP^{(g)}_R\{ p\} =  {\rm Schur}_{R}\{p\}
+ \sum_{{R'< R}} {\cal K}_{{R,R'}}^{(g)} \cdot {\rm Schur}_{R'}\{p\}
\\ \\
\widehat{\SchurP}_R^{(g )}\{p\}
=\Schur_{R}\{p\}\ + \  \sum_{ R'^\vee> R^\vee}
\widehat{\cal K}^{(g )}_{R'^\vee,R^\vee}\cdot\Schur_{R'}\{p\}
\end{array}
\right\}
}
\label{PthroughSchurs}
\label{duPthroughSchurs}
\ee
where the Young diagrams are ordered lexicographically:
\be
R>R'  \ \ {\rm if} \ \ r_1>r_1' \ \ {\rm  or\ if} \ \ r_1=r_1', \ {\rm but} \ r_2>r_2',
\ \ {\rm or\  if} \ \ r_1=r_1' \ {\rm  and} \ r_2=r_2', \ {\rm but} \ r_3>r_3',
\ \ {\rm and\ so\ on}
\label{lexico}
\ee
The pair of functions emerges because this ordering is not consistent
with the transposition of Young diagrams:
\be
R>R'\ \text{is not always the same as} \ {R'}^\vee >R^\vee
\label{transpolex}
\ee
The first discrepancy appears at level $|R|=6$, it is the pair
$[3,1,1,1]>[2,2,2]$, for which $[3,1,1,1]^\vee = [4,1,1]>[2,2,2]^\vee=[3,3]$.
}

\item{
A similar definition through the triangular transformation is also most efficient
for the generalized Macdonald functions \cite{genMac},
which depend on several Young diagrams and several sets of time-variables.
Moreover, it seems also applicable to the plane partition generalizations \cite{Morlast}.
In the conventional approach, in all these cases 
the triangularity is associated with a peculiar (triangular) choice of the Hamiltonian deformations,
which appear relevant to the Calogero-Ruijenaars systems and to matrix models
(and thus to the AGT relations).
One of the possible explanations for this can be through the PQ (spectral) duality, 
which is the basic semi-hidden symmetry of generic network models.
}

\item{
The Macdonald-Kostka coefficients ${\cal K}_{R,R'}^{(g)}$
can be defined iteratively in $R$ and $R'$
from the orthogonality conditions
\be
\Big<\SchurP^{(g)}_R\Big|\SchurP^{(g)}_{R'}\Big>
= ||\SchurP^{(g)}_R||^2 \cdot \delta_{R,R'}
\label{orthoc}
\ee
with the scalar product given by
\be
\Big< { p}^{\Delta}\Big| { p}^{\Delta'} \Big>^{(g)} =
z_\Delta \cdot \delta_{\Delta,\Delta'}\cdot
\left(\prod_{i=1}^{l_\Delta} g_{\delta_i}\right)
\label{scapro}
\ee
Here the Young diagram $\Delta=\big[\delta_1\geq\delta_2\geq\ldots\geq \delta_{l_\Delta}\big]$,
and $p^\Delta = \prod_{i=1}^{l_\Delta} p_{\delta_i}$.
The combinatorial factor $z_\Delta$ is best defined in the dual parametrization of the
Young diagram, $\Delta = \big[\ldots,2^{m_2},1^{m_1}\big]$, then
$z_\Delta = \prod_k k^{m_k}\cdot m_k!$.
Note that the normalization of $\SchurP^{(g)}$ is already fixed by  the choice of
unit diagonal coefficient (the first term) in (\ref{PthroughSchurs}):
\be
{\cal K}^{(g)}_{RR}=1
\ee
therefore the norm $||\SchurP^{(g)}||$ is a deducible quantity.
}

\item{
The Schur functions $\Schur_R=\SchurP^{(g=1)}_R$ are orthonormal at $g=1$:
\be
\underline{
\Big<\Schur_R\Big|\Schur_{R'}\Big>^{(g=1)} = \delta_{R,R'}
}
\label{Schurscapro}
\ee
The Macdonald-Kostka matrices ${\cal K}^{(g)}$ and $\widehat{{\cal K}}^{(g)}$
(both!) diagonalize their product matrix
with $g\neq 1$,
\be
\mu_{R,R'}=\Big<\Schur_R\Big|\Schur_{R'}\Big>^{(g)}
=\sum_\Delta \ {\psi_{R_1}(\Delta)\psi_{R_2}(\Delta)\over z_\Delta}\cdot g_\Delta
\label{Schurscaprog}
\ee
i.e. provide its Gauss decomposition for two different
choices of the triangular (Borel) structure.
Here $\psi_R(\Delta)$ are the symmetric group characters.
Note that $g_\Delta = \prod_{i=1}^{l_\Delta} g_{\delta_i}$ appear here in a combination,
similar to $p^\Delta$ in (\ref{scapro}), and this is the first manifestation of
a far-going similarity between $g_n$ and time-variables, which we will further
review in detail in sec.\ref{gvars} below.

The Gauss decomposition is a very efficient operation in computer simulations.
An additional advantage of (\ref{Schurscaprog}) is that the orthogonality of characters
$\psi_R(\Delta)$ implies an equally simple
expression for an arbitrary power of $\mu$ including the inverse matrix $\mu^{-1}$,
the one which should actually be decomposed to provide the Kostka matrix ${\cal K}^{(g)}$:
\be
\Big(\mu^{-1}\Big)_{R,R'}=\sum_\Delta \ {\psi_{R_1}(\Delta)\psi_{R_2}(\Delta)\over z_\Delta}\cdot {1\over g_\Delta}=
\mu^{(g^{-1})}_{R,R'}\ \ \ \ \ \Longleftarrow \ \ \ \ \
\Big(\mu^{n}\Big)_{R,R'}=\mu^{(g^{n})}_{R,R'}\ \ \ \ \ \ n\in\mathbb{Z}
\ee

The orthogonality conditions  (\ref{orthoc})
can also be used to represent the Macdonald-Kostka coefficients ${\cal K}_{R,R'}^{(g)}$
in (\ref{PthroughSchurs})  as
determinants/minors of the matrix $\mu_{R,R'}$.
}

\item{
Being a complete basis in the space of $p$-polynomials,
the Kerov functions satisfy the
multiplication rules
\be
\boxed{
\begin{array}{c}
\\
\SchurP^{(g)}_R\{p\} \cdot \SchurP^{(g)}_{R'}\{p\} =\displaystyle{
\sum_{R\cup R'\leq R''\leq R+R'}} N^{R''}_{R,R'}(g)\cdot \SchurP^{(g)}_{R''}\{p\}
\\  \\
\widehat{\SchurP}^{(g)}_R\{p\} \cdot \widehat{\SchurP}^{(g)}_{R'}\{p\} =
\displaystyle{\sum_{R^\vee+R'^\vee\leq R''^\vee\leq R^\vee\cup R'^\vee }}
\widehat{N}^{R''}_{R,R'}(g)\cdot \widehat{\SchurP}^{(g)}_{R''}\{p\}
\end{array}}
\label{Kerprod}
\ee
and the sums are rather restricted, since the generalized Littlewood-Richardson coefficients $N^{R''}_{R,R'}(g)$ are non-zero only
in between the partitions $R\cup R' = [r_1+r'_1,r_2+r_2',\ldots]$ and $R+R' = [{\rm ordered\ collection\ of\ all}\
r_i \ {\rm and} \ r_j']$.
}

\item{
Also obviously existing is the decomposition
\be
\SchurP_R^{(g)}\{p+p'\} = \sum_{R'\subset R} \SchurP_{R'}^{(g)}\{p\}
\cdot \SchurP_{R/R'}^{(g)}\{p'\}
\label{Kersum}
\ee
which defines the {\it skew} Kerov functions, and it is actually a dual of (\ref{Kerprod}),
because (not quite trivially)
\be
\boxed{
\SchurP_{R/R'}^{(g)}\{p\} =
\sum_{R''} \widehat{N}^{R^\vee}_{{R'}^\vee {R''}^\vee}(g^{-1}) \cdot
\SchurP_{R''}^{(g)}\{p\}
}
\ee
with the same $N$ as in (\ref{Kerprod}).
}

\item{
Equivalence of the two latter decompositions is consistent with the transposition rule
\be
\boxed{
\begin{array}{c}
\SchurP_{R }^{(g)}\{p\} =
(-)^{|R|}\cdot{||\SchurP_{R }^{(g)}||^2 }\cdot
\widehat{\SchurP}^{(g^{-1})}_{R^\vee}\{ -p^\vee  \}
\label{Ptransp}
\\ \\ \Updownarrow \\ \\
\SchurP_{R }^{(g)}\{p^\wedge\} =
(-)^{|R|}\cdot{||\SchurP_{R }^{(g)}||^2 }\cdot
\widehat{\SchurP}^{(g^{-1})}_{R^\vee}\{ -p   \}
\end{array}
}
\label{Ptransp}
\ee
where operation $^\wedge$ on the time-variables is inverse of $^\vee$:
$(p^\wedge)^\vee = p$, and, explicitly,
\be
p^\vee_k = \frac{p_k}{g_k} \ \ \ \ {\rm and} \ \ \ \ p^\wedge_k =  g_k\cdot p_k
\label{ transpochange}
\ee
}

\item One can also introduce a dual Kerov function (in complete analogy with the Macdonald case, \cite{Mac}):
\be
\boxed{
\underline{\SchurP}_{R }^{(g)}\{p\} :=
{\SchurP_{R }^{(g)}\{p\}\over ||\SchurP_{R }^{(g)}||^2 }}
\ee
With this definition, formula (\ref{Ptransp}) looks a bit simpler
\be
\underline{\SchurP}_{R }^{(g)}\{p\} =
(-)^{|R|}
\widehat{\SchurP}^{(g^{-1})}_{R^\vee}\{ -p^\vee  \}
\ee
Moreover, introducing the corresponding dual skew Kerov functions,
\be
\underline{\SchurP}_R^{(g)}\{p+p'\} = \sum_{R'\subset R} \underline{\SchurP}_{R'}^{(g)}\{p\}
\cdot \underline{\SchurP}_{R/R'}^{(g)}\{p'\}
\label{Kersum}
\ee
one obtains
\be
\underline{\SchurP}_{R/R'}^{(g)}\{p\} =
\sum_{R''} {N}^{R}_{{R'} {R''}}(g) \cdot
\underline{\SchurP}_{R''}^{(g)}\{p\}
\ee

\item{
Important in applications is the
Cauchy summation formula:
\be
\boxed{
\sum_{R}(-)^{|R|}\SchurP_R^{(g)}\{p\}\cdot\widehat{\SchurP}^{(g^{-1})}_{R^\vee}\{-{p'}\}
=\sum_R \frac{\SchurP_R^{(g)}\{p \}\cdot\SchurP_R^{(g)}\{p'^\wedge\}}{||\SchurP_R^{(g)}||^2}
= \exp\left( \sum_k \frac{p_kp_{k}'}{k}\right)
}
\label{Cauchy}
\ee
}
and, more generally,
\be
\sum_{R}(-)^{|R|}\SchurP_{R/\eta}^{(g)}\{p\}\cdot
\widehat{\SchurP}^{(g^{-1})}_{R^\vee/\zeta}\{-{p'}\}
= \exp\left( \sum_k \frac{p_kp_{k}'}{k}\right)
\cdot
\sum_\sigma (-)^{|\zeta|+|\eta|+|\sigma|}\SchurP_{\zeta^\vee/\sigma}^{(g)}\{p\}\cdot
\widehat{\SchurP}^{(g^{-1})}_{\eta^\vee/\sigma^\vee}\{-{p'}\}
\ee
where the sum over $\sigma$ at the r.h.s. contains only finitely many terms.

\end{itemize}

\section{Specialization of the function $g$}

\begin{itemize}

\item{
In the FZ parametrization \cite{FZ} motivated by applications to the XYZ model (at $\kappa=2$), the function $g_n$ depends on several $q$ and $t$ parameters:
\be
g_n = \prod_{\alpha=1}^{\kappa} \frac{\{q_\alpha^n\}}{\{t_\alpha^n\}}
\label{Kerovpar}
\ee
where $\{x\} = x-x^{-1}$.
For the Schur functions, $\underline{\kappa=0}$,
for the Macdonald functions, $\underline{\underline{\kappa=1}}$.
Thus, the Schur functions are orthogonal in the product
\be
\underline{
\Big< { p}^{\Delta}\Big| { p}^{\Delta'} \Big>^{(1)} =
z_\Delta \cdot \delta_{\Delta,\Delta'}
}
\ee
and their norm is $\underline{||\SchurP_R^{(\kappa=0)}||^2=1}$,
while the Macdonald functions $\Mac\{q,t,p\}$, are orthogonal in
\be
\underline{\underline{
\Big< { p}^{\Delta}\Big| { p}^{\Delta'} \Big>^{({\rm Mac})} =
z_\Delta \cdot \delta_{\Delta,\Delta'}\cdot \prod_{i=1}^{l_\Delta}
\frac{\{q^\delta_i\}}{\{t^{\delta_i}\}}
}}
\ee
and their norm is
\be
\underline{\underline{
||\SchurP_R^{(\kappa=1)}||^2={h_{R^\vee}(t,q)\over h_R(q,t)}
}}
\ee
where
\be
h_R(q,t):=\prod_{i,j\in R} (q^{j-R_i}t^{i-R^\vee_j+1}-q^{R_i-j}t^{R^\vee_j-i-1})
\ee
}

\item{
If $g_n$ are considered as additional time-variables, Macdonald locus
$\underline{\underline{g_n = \frac{\{q^n\}}{\{t^n\}} }}$
becomes a direct counterpart of the topological locus \cite{DMMSS}
$p_k=\frac{\{A^k\}}{\{t^k\}}$ for the ordinary time-variables.
This analogy nicely explains both the distinguished role of Macdonald case
in many circumstances, as well as its limitations and the need of
further extension to the full-fledged Kerov case with arbitrary $g_n$.
The main peculiarity of topological locus are nice factorization properties,
but in the case of, say, knot theory, they are directly helpful in evolution formulas \cite{evo}
and in differential expansions \cite{diffexpan},
while some equally important properties like integrability \cite{MMM1}
in {\it torus} links and knots \cite{RJ,DMMSS,AgSha,Che},
are available only if one looks from beyond the topological locus \cite{MMM2}.
What really matters is {\it superintegrability}, it is {\it reflected}
in factorization at the topological/Macdonald loci, but is actually a far
more general property of the Schur functions  and their relatives,
which holds far beyond these small 2-dimensional varieties in the
infinite-dimensional space of time-variables,
and is, perhaps, extendable to the entire world of Kerov functions.
As we suggest in sec.\ref{gvars} below, it can be not only extendable,
but becomes much richer after this extension is done:
the symmetries are not really lost, but rather enhanced.
}

\end{itemize}

\section{Relation to symmetric polynomials and to representation theory }

\begin{itemize}

\item{
On the $n$-dimensional Miwa locus in the time-variable space,
\be
{p}_k=\tr X^k = \sum_{a=1}^n x_a^n
\ee
the function $\SchurP_R$ turns
into a symmetric polynomial
\be
\SchurP_R[X] = \sum_{R'<R} k^{(g)}_{R,R'}\cdot m_{R'}[X]
\label{Schurviam}
\ee
which is a triangular sum of monomial symmetric polynomials
\be
m_{R}[X] = {\rm symm}_{x} \left(\prod_{a=1}^{l_R} {x_a}^{r_a}\right)
\ee
}

\item{
The Kostka numbers
\be
\underline{
k_{R,R'}^{(\Schur)} = {K^{-1}_{R,R'}}({q=0,t=1})
}
\ee
are the values of  inverse
Macdonald-Kostka matrix $K$ at the particular values of parameters $q=0$ and $t=1$
(inverted is the matrix, not the particular entries).
This is the original Macdonald's requirement that, at these values, his polynomials
coincide with $m_R$: $\ M_R[0,1,X] = m_R[X]$.
}

\item{
Conceptually, the Schur functions are symmetric polynomials associated with
representation theory of $GL_n$
(\underline{\underline{generalizations to other root systems also exist}}).
They are defined so that ${\rm Schur}[X]$ for $n={\rm rank}(GL_{n})$
are the characters (elements of the center of the group algebra) of $GL_n$.
As a corollary,
the sum in multiplication rule (\ref{Kerprod}) is now restricted to
\be
\underline{\underline{R''\in R\otimes R'}}
\label{selruleforN}
\ee
In the Kerov case, the sum includes contributions not appearing in the product
of representations.
For the first time, this happens at the level $|R|$, for example,
\be
\SchurP_{[4]}\cdot \SchurP_{[1,1]} =
\alpha\cdot \SchurP_{[4,1,1]}+\beta\cdot\SchurP_{[4,2]}+\gamma\cdot\SchurP_{[5,1]}
\label{4prod11}
\ee
while $[4]\otimes [1,1] = [4,1,1]+[5,1]$.
Moreover, for the $\widehat K$-Kerov functions, the same relation contains
not three, but four terms:
\be
{\SchurP}_{[4]}\cdot \SchurP_{[1,1]} =
\hat\alpha\cdot \widehat{\SchurP}_{[4,1,1]}+\hat\beta\cdot\SchurP_{[4,2]}
+\hat\delta\cdot\widehat{\SchurP}_{[3,3]}
+\hat\gamma\cdot\SchurP_{[5,1]}
\label{hat4prod11}
\ee
because there are two diagrams $[2,2,1,1]=[4,2]^\vee$ and $[2,2,2]=[3,3]^\vee$
between $[2,1,1,1,1]=[5,1]^\vee$ and $[3,1,1,1]=[4,1,1]^\vee$.
We made it explicit
that only two of the four emerging $\widehat K$-Kerov functions
are different from the $K$-ones.
}

\item{
In the Macdonald case, however, (\ref{selruleforN}) is believed to survive.
In particular, the coefficient $\beta$ in (\ref{4prod11}),
vanishes in the Schur and Macdonald cases $\kappa=0,1$, but not for $\kappa>1$.

As a related (but not equivalent) property, on the three-dimensional locus
$g(k)=x_1^k+x_2^k+x_3^k$
\be
\beta \sim
x_1^2x_2^2x_3^2(x_1^2-x_2x_3)(x_2^2-x_1x_3)(x_3^2-x_1x_2) \cdot B(x)
\label{beta3loc}
\ee
with a complicated irreducible polynomial $B(x)$.
There is also a relatively simple denominator.
}

\end{itemize}

\section{Equivalence of $\widehat{\SchurP}$ and $\SchurP$ in Macdonald case}

Actually only half of the Macdonald  functions needs to be calculated directly,
the other half can be obtained from (\ref{Ptransp}).
This is a big calculational advantage, because the computer time needed for
orthogonalization grows dramatically
with the number of terms in the sum (\ref{PthroughSchurs}),
and reducing the problem to sums of twice smaller length allows one to calculate the
Kerov/Macdonald functions for higher levels $|R|$.

At the same, for the Kerov functions this is not quite correct because of the two types of the Kerov functions related by (\ref{Ptransp}). In other words, (\ref{Ptransp}) is a duality, and not self-duality condition in the generic Kerov case.

\begin{itemize}

\item{
It is remarkable that (\ref{Ptransp}) is a self-duality condition not only in the Schur,
but also in the Macdonald case
of $\kappa=1$ when it is sufficient to define
$\underline{\underline{\widehat{\Mac} = \Mac}}$
so that (\ref{Ptransp}) becomes
\be
\underline{\underline{
\Mac_{R}\{q,t,\,p_k\} =
(-)^{|R|}\cdot ||\Mac_{R }||^2\cdot
\Mac_{R^\vee }\left\{t,q,\, -\frac{\{t^k\}}{\{q^k\}}\,p_k\right\}
}}
\label{Mactrans}
\ee
In the Schur case, this further simplifies to just
\be
\Schur_{R^\vee}\{p\} = (-)^{|R|}\cdot \Schur_R\{-p\}
\ee
}

\item{
The reason for this simplification is that, at $\kappa=1$, all the  problematic Macdonald-Kostka coefficients,
i.e. those for the pairs of diagrams, when both $R>R'$ and $R^\vee>R'$, are vanishing,
for example
\be
\underline{\underline{
{\cal K}_{[3,1,1,1],[2,2,2]}^{({\rm Mac})}=0
}}
\label{lev6domK}
\ee
This vanishing condition imposes severe restrictions on the function $g$,
which are satisfied for $\kappa=1$, i.e. for the Macdonald deformation,
but are violated for $\kappa>1$, i.e. for the generic Kerov functions.

The analogue of (\ref{beta3loc}) is also true:
on the 3-dimensional locus $g(k)=x_1^k+x_2^k+x_3^k$ this Macdonald-Kostka coefficient reduces to
\be
{\cal K}_{[3,1,1,1],[2,2,2]} \sim
x_1^2x_2^2x_3^2(x_1^2-x_2x_3)(x_2^2-x_1x_3)(x_3^2-x_1x_2) \cdot \Schur_{[7]}\{g\}
\label{kos3loc}
\ee
but the substitute of the polynomial $B(x)$ in this case is actually much simpler. Surprisingly, one again arrives at the Macdonald properties of the group theory expansion (\ref{selruleforN}) and self-duality (\ref{Mactrans}) on the two-dimensional locus, where both
$\beta$ in (\ref{beta3loc}) and ${\cal K}_{[3,1,1,1],[2,2,2]}$ in (\ref{kos3loc}) vanish. It can be chosen, for instance, at $x_3^2=x_1x_2$.
}

\item{
Actually, in the Schur and Macdonald cases,
one can use in   (\ref{PthroughSchurs})
any {\it partial} ordering of
Young diagrams which satisfies the {\it dominance rule}:
\be
\underline{\underline{
R\geq R' \ \ {\rm if} \ \  \sum_{a=1}^i {r_a} \geq \sum_{a=1}^i {r_a'}
\ \ {\rm for\ all} \ \ i
}}
\label{domrule}
\ee
since, for all pairs which remain unordered, the corresponding Kostka numbers vanish,
together with their inverse, e.g.
\be
\underline{\underline{
k^{\pm 1}_{[3,1,1,1],[2,2,2]}=0
}}
\label{lev6dom}
\ee
(inversion here, as in similar cases before, is inversion of the entire triangular matrix, not
of the particular entry).
}

\end{itemize}

\noindent
To finish this piece of the story, we emphasize once again
that the  duality (\ref{Ptransp}) is rather non-trivial, and
it guarantees that the Kerov/Macdonald functions for symmetric representations
and "nearly-symmetric" are actually as simple as those for the
antisymmetric, i.e. equivalence of the symmetric and antisymmetric "ends"
of the partition list,
which is far from obvious from the definition (\ref{PthroughSchurs}),
where the antisymmetric functions coincide with the antisymmetric Schur polynomials,
while symmetric are combinations of all Schur functions at a given level.

\section{Properties peculiar for the Schur functions}

\begin{itemize}

\item{
Specific for the Schur functions {\it per se} is determinant expression
through those in symmetric representations
\be
\underline{
S_R = {\det}_{i,j=1}^{l_R} S_{[r_i-i+j]},\ \ \ \ \ \ S_{[<0]}=0
}
\label{detSchur}
\ee
Deformations of this formula exist,
for example, one can treat in this way the
determinant formula for the Macdonald-Kostka coefficients.

}

\item{The
Schur functions are expanded in symmetric-group characters $\psi_R(\Delta)$:
\be
\underline{
\Schur_R\{p\} = \sum_{\Delta}  \frac{\psi_R(\Delta)}{z_\Delta}\cdot p^\Delta
}
\label{Schurtnroughpsi}
\ee
and their orthogonality (\ref{Schurscapro}) is related to the orthogonality of characters,
\be
\underline{
\sum_\Delta \frac{\psi_R(\Delta)\psi_{R'}(\Delta)}{z_\Delta} = \delta_{R,R'}
}
\ee
In both these formulas, the sums are over Young diagrams of size $|\Delta|=|R|=|R'|$.
Possible Kerov and Macdonald deformations of (\ref{Schurtnroughpsi})
would require a deep understanding
of the Schur-Weyl-Howe duality \cite{Howe}
and is related \cite{DIMDAHA} to deep and yet unanswered questions about the
DIM algebra \cite{DIM} and its further generalizations.
}

\item{

As described in detail in \cite{MMN},
the Schur functions are the common eigenfunctions of the infinite set of commuting generalized
cut-and-join
operators $\hat W_\Delta$
with eigenvalues made from symmetric group characters $\psi_R(\Delta)$:
\be
\underline{
\hat W_\Delta\left\{p_k,\frac{\p}{\p p_k}\right\} \,\Schur_R\{p\} =
\frac{\psi_R(\Delta)}{d_Rz_\Delta}\cdot \Schur_R\{p\}
}
\label{Wops}
\ee
where $d_R:=\Schur_R\{\delta_{k,1}\}$.
Existence of such operators at each level is a trivial linear-algebra property
easily generalizable to the Macdonald
and Kerov functions (as well as far beyond them, see, for example, \cite{Morlast}).
However, in the Schur (and, partly, Macdonald) case, they can be "unified" into
an operator made by the Sugawara construction \cite{Sug} from a linear Cherednik-Dunkl operator
$\hat J$ (a counterpart of the Kac-Moody current in the WZW model \cite{WZW}):
$\underline{\underline{\hat W_R \sim \Tr \hat J^R}}$, which
interrelates naive operators at different levels.
Getting a deeper understanding of these relations and their proper extension beyond the Schur case
is one of the many open questions in the field.

}

\item{
The Gaussian averages of the Schur functions provide the same functions
evaluated at the topological locus \cite{MMchar}:
\be
\underline{
\int_{n\times n} \Schur_R[X] \ e^{-\tr X^2} dX \sim \Schur_R\{p_k=N\} =\hbox{dim}_N(R)
}
\nn \\
\underline{
\int_{n\times n} \Schur_R[e^X] \ e^{-\frac{1}{\hbar}\tr X^2} dX \sim
\Schur_R\left\{p_k=\frac{\{e^{\hbar N}\}}{\{e^\hbar\}}\right\} =D_N(R)
}
\label{avers}
\ee
which are ordinary and quantum dimensions of the representation $R$ of $SL_n$
and $U_{e^\hbar}(SL_n)$ correspondingly.
This property is responsible for {\it the superintegrability} of matrix models \cite{SIMaMo},
it has important generalizations in many directions, from knot theory to
a non-trivial extension to tensor models \cite{IMMlast}.
Its Macdonald generalization is straightforward \cite{MPSh},
but it involves the Jackson integrals, which have no yet a straightforward
generalization to the generic Kerov deformations.
}

\item{
One could think that (\ref{detSchur})-(\ref{avers})
are the only properties of the Schur functions
which get significantly changed (into much "heavier" formulas)
under the Kerov/Macdonald  deformation, but we have found another one.
As already mentioned at the beginning of this paper,
the definition of Schur {\bf functions in the composite representations}
also appears to be deformed not straightforwardly.
An explanation of this fact and the ways to overcome the problem
will be a starting point of our far-more-speculative
consideration in the next paper \cite{compomac}.
}

\end{itemize}

\section{Measure $g$-variables as the new times
\label{gvars}}

As will be illustrated by examples in the Appendix,
values of the function $g_n$ can actually be treated as new time-variables so that the Kerov functions can be efficiently expressed
through the Schur functions of these variables.
Moreover, this expansion is somewhat different for the diagrams, which appear
at the beginning and at the end of the lexicographical ordering,
and the relation between the two expansions is not {\it fully} exhausted by
(\ref{Ptransp}): there are additional symmetries between the structures for
the $k$-th diagram at the beginning and the $k-1$-th one at the end.
Already from (\ref{Ptransp}) is follows that if the Schur functions of $p_k$ and $g_k$
appear in one expansion, they become functions of $p^\vee_k=p_k/g_k$
in another, and it appears that $g_k$ is also substituted by $g_k^{-1}$.
The peculiarity of this part of the story is that the expansion in $p_k^\vee$
is much simpler than the one in $p_k$, which is originally used in (\ref{PthroughSchurs}).
This expansion is upper instead of lower triangle and, in the Macdonald case,
its coefficients are known to be rather simple minors  \cite{LLM}.
It looks like
this result is extendable to the Kerov functions:
the coefficients can be expressed through the Schur functions at $p_k=g_k^{-1}$.

Note that, while $g_k$ can clearly be treated as new (additional) time-variables,
they participate in additional operations like inversion and multiplication,
which are unusual for time-variables.
In particular, these operations are not seen on the Miwa loci,
and thus are not captured by the usual symmetric functions technique;
this can be the reason why these interesting structures were not noticed,
and thus theory of the Kerov functions remained underestimated and nearly undeveloped.

\begin{itemize}

\item{
The Kerov functions are polynomials in $p$, but as functions of $g$ they are rational,
with $g$-dependent denominators, which are related through (\ref{Ptransp})
to the norms of Kerov functions for transposed diagrams,
\be
\Ker^{(g)}_R\{p\} = \frac{{\rm pol}_R(p,g)}{\Delta_R\{g\}}
\ee
}

\item{
For the first diagram in the lexicographical ordering, the Kerov function is independent of $g$,
and the denominator is unity:
\be
\Ker^{(g)}_{[1^r]}\{p\} = \Schur_{[1^r]}\{p\}, \ \ \ \ \
\boxed{ \Delta_r^{(1)}:=\Delta_{[1^r]} = 1 }
\ee
}

\item{
However, already for the second diagram $[2,1^{r-2}]$
(associated with the adjoint representation of $SL_r$) there is
a non-trivial Macdonald-Kostka coefficient, which is actually
the ratio of two Schur functions of $g$-variables:
\be
\boxed{
{\cal K}_{[2,1^{r-2}],[1^r]}^{(g)} = -\frac{ \Schur_{[r-1,1]}\{g\}}{\Schur_{[r]}\{g\}}
}
\label{firstKostka}
\ee
This implies that
\be
\Ker_{[2,1^{r-2}]}^{(g)}\{p\} = \Schur_{[2,1^{r-2}]}\{p\} -
\frac{\Schur_{[r-1,1]}\{g\}}{\Schur_{[r]}\{g\}}\cdot \Schur_{[1^r]}\{p\}
\label{Keradj}
\ee

Thus, the next denominator is
\be
\boxed{
\Delta^{(2)}_r := \Delta_{[2,1^{r-2}]} =   \Schur_{[r]}\{g\}
}
\ee
}
If multiplied by $r!$  it is a polynomial with integer coefficients.

\item{
In the case of the third diagram $[2,2,1^{r-2}]$, the denominator is described
by a somewhat strange 3-term expression:
\be
\Delta^{(3)}_{r} := \Delta_{[2,2,1^{r-2}]} 
=  \det\left(\begin{array}{ccc}
\Schur_{[r]} & 0 & \Schur_{[1]} \\
\Schur_{[r]} & \Schur_{[r-1]} & 0 \\
\Schur_{[r-1]}  & \Schur_{[r-2]} & 1
\end{array} \right)
\label{Delta3}
\ee
This time it becomes an integer polynomial after multiplication by   $r!(r-2)!$.
It appears in similarly looking formulas for the Macdonald-Kostka coefficients
\be
\boxed{
\begin{array}{c}
\\
{\cal K}_{[2,2,1^{r-4}],[1^r]}^{(g)} = -
\frac{\Schur_{[r-1,1]}\cdot\Schur_{[r-1]}
- \Schur_{[r-1,1]}\cdot\Schur_{[r-3]}\cdot\Schur_{[2]}
+ \Schur_{[r-2,2]}\cdot\Schur_{[r-2]}\cdot\Schur_{[1]}}
{ \Delta^{(3)}_{r} }
 \\ \\
{\cal K}_{[2,2,1^{r-4}],[2,1^{r-2}] }^{(g)} =-
\frac{ \Schur_{[r-1,1]}\cdot\Schur_{[r-1]}
- \Schur_{[r-1,1]}\cdot\Schur_{[r-3]}\cdot\Schur_{[2]}
+ \Schur_{[r-3,1]}\cdot\Schur_{[r-1]}\cdot\Schur_{[2]} }{ \Delta^{(3)}_{r} }
\\ \\
\end{array}
}
\label{Kostka2}
\ee
and, as a corollary, for the third Kerov function $\Ker^{(g)}_{[2,2,1^{r-4}]}$.
Note that the  two functions in (\ref{Kostka2})
differ only in their last terms, the first two are the same.
}

\item{
The truly interesting is the fourth diagram, $[2,2,2,1^{r-6}]$, because it is the first one
where the two orderings become different:
$[2,2,2,1^{r-6}]<[3,1^{r-6}]$ but also
$[2,2,2,1^{r-6}]^\vee=[r-3,3]<[r-2,1,1] =[3,1^{r-6}]^\vee$;
in other words, $[2,2,2,1^{r-6}]$ is the fourth in the lexicographical ordering,
but in transposed ordering the fourth diagram is  $[3,1^{r-6}]$.
To get an insight of what to expect from $\Delta^{(4)}$
relevant in this case,
it is practical to consider a complementary part of the story,
what we will do next.
We return to $\Delta^{(4)}$ in (\ref{Delta4}) below.
}

\item{
As one can anticipate from (\ref{Ptransp}), the Kerov function in
symmetric representation, which is the last in the lexicographical ordering,
is almost as simple as in the antisymmetric one, which is the first,
despite this is not easy to see directly from the definition (\ref{PthroughSchurs}).
In fact, (\ref{Ptransp}) implies that $\Ker_{[r]}$ is a Schur polynomial of
$p_k/g_k$, but the $g$-dependence of the prefactor still needs to be found.
In this case, it is easy.
According to (\ref{Ptransp}), what is needed is the norm
of the antisymmetric Schur function:
\be
\Ker^{(g)}_{[r]}\{p\} \ \stackrel{(\ref{Ptransp})}{=} \
\frac{ (-)^{|R|}\,\Ker_{[1^r]}^{(g^{-1})}\{-p_k/g_k\}} {||\Ker_{[1^r]}^{(g^{-1})}||^2}
=  \frac{ (-)^{|R|}\,\Schur_{[1^r]}\{-p_k/g_k\}} {||\Ker_{[1^r]}^{(g^{-1})}||^2}
=  \frac{ \Schur_{[r]}\{p_k/g_k\}} {||\Ker_{[1^r]}^{(g^{-1})}||^2}
\ \ \Longrightarrow
\nn \\
\Delta^\vee_r \sim ||\Ker^{(g^{-1})}_{[1^r]}||^2
= \Big< \Schur_{[1^r]}\Big| \Schur_{[1^r]}\Big>^{(g^{-1})} \
\stackrel{(\ref{Schurtnroughpsi})+(\ref{scapro})}{=} \
\sum_{\Delta\vdash r} \frac{\psi_{[1^r]}(\Delta)^2}{z_\Delta}\frac{1}{g^\Delta}
=\Schur_{[r]}\{g^{-1}_k\}
\ee
where $g^\Delta = \prod g_{\delta_i}$, and, at the last stage, we used peculiarities
of antisymmetric and symmetric representations:
$\psi_{[1^r]}(\Delta)=(-)^{l_R+|R|}$ and $\psi_{[r]}(\Delta)=1$
for any $\Delta$ of the size $r$. $l_R$
Thus, for the Kerov functions in symmetric representations, we get a universal expression
\be
\boxed{
\Ker_{[r]}^{(g)}\{p\} =
\frac{\Schur_{[r]}\left\{\frac{p_k}{g_k}\right\}}{\Schur_{[r]}\left\{\frac{1}{g_k}\right\}}
}
\label{Kersym}
\ee
explicitly realizing their relation (\ref{Ptransp}) to the $g$-independent Kerov functions
in antisymmetric representations
\be
\boxed{
\Ker_{[1^r]}^{(g)}\{p\} = \Schur_{[1^r]}\{p_k\}
\label{Kerantisym}
}
\ee
}

\item{
Remarkably, the denominator in (\ref{Kersym}), which has no counterpart in
(\ref{Kerantisym}), is related to the one in (\ref{Keradj}):
\be
\boxed{
\Delta_{[r]}\{g_k\} = \Delta_{r}^{(2)}\{g_k^{-1}\}
}
\ee
}

\item{
Moreover, this property persists further: the
denominator in the case of penultimate (second from the end) diagram
$[r-1,1]$ is obtained by $g$-inversion from that for the third one, $[2,2,1^{r-2}]$:
\be
\Delta_{[r-1,1]}\{g_k\} = \Delta_r^{(3)}\{g_k^{-1}\}
\ee
Given (\ref{Delta3}), this is an absolutely explicit formula,
and it enters  the equally explicit expression for the corresponding Kerov function:
\be
\boxed{
\Ker^{(g)}_{[r-1,1]}\{p\} =
\frac{
\Schur_{[r]}\left\{\frac{1}{g_k}\right\}\cdot\Schur_{[r-1,1]}\left\{\frac{p_k}{g_k}\right\} -
{\Schur_{[r-1,1]}\left\{\frac{1}{g_k}\right\}}\cdot
\Schur_{[r]}\left\{\frac{p_k}{g_k}\right\}}
{\Big(\Schur_{[1]}\cdot\Delta^{(3)}_r\Big)\left\{\frac{1}{g_k}\right\}}
}
\ee
Note that, while the denominator is related to that for $\Ker^{(g)}_{[2,2,1^{r-4}]}$
(this is not quite the relation (\ref{Ptransp}), the two diagrams are not transposed,
but additionally shifted by one in the lexicographical ordering!),
this is not immediately so for the numerators.
Instead, the numerators of the Kerov functions in the vicinity of symmetric case
demonstrate a structure of minor expansion, which should be a straightforward generalization
of the one found in \cite{LLM} for the Macdonald functions.
}

\item{
Since for the particular value of $r=3$, the first diagram from the end coincides with the
third one from the beginning, we can expect that, in this case, $\Delta^{(2)}\{g^{-1}\}$
when expanded in $g$ will have the same shape as $\Delta^{(3)}$, i.e.
the Schur function of inverted $g$ will be a three-term determinant.
This is indeed the case, and it implies that the shape of $\Delta_{[r-2,2]}$
can actually reveal the structure of $\Delta^{(4)}\{g^{-1}\}$ and
thus of $\Delta^{(4)}\{g\}$.
Already  the simplest $\Delta_{[2,2]}$ is given by a somewhat sophisticated
19-term formula (see the very end of the next section),
and this illustrates the type of problems one should deal with and resolve.
}

\item{
The full expression for $\Delta^{(4)}$ is rather lengthy,
for illustrative purposes, we present just the first three terms,
which fully capture the contribution with the product $g_{[r]}g_{[r-1]}$:
\be
\left.
\begin{array}{c}
\Delta^{(4)}_r := \Delta_{[2,2,2,1^{r-6}]} \\ \\
\widehat{\Delta}^{(4)}_r :=  \widehat{\Delta}_{[3,1^{r-3}]}
\end{array} \right\} =
\ \ \ \ \ \ \ \ \ \ \ \ \ \  \ \ \ \ \ \ \ \ \ \ \ \ \ \ \ \ \ \ \ \ \ \ \ \ \ \ \ \
\nn \\
=
\Schur_{[r]}\cdot \Schur_{[r-1]} \cdot\Big(\Schur_{[r-2]}\Schur_{[2]}
+\Schur_{[r-3]}\Schur_{[1]}^3+\Schur_{[r-4]}\Schur_{[2]}^2\Big) + \ldots
\label{Delta4}
\ee
(for example of the full expression in a particular case, see the last formula
in (\ref{Deltalist}) in the Appendix).
This time the factorial would be $((r-2)!)^2(r-4)!$.
If so normalized, $\Delta^{(4)}_r$ at
even $r$  are divisible by $2$ (i.e. are even integer polynomials).
What is important, the hatted and ordinary quantities  are the same,
\be
\boxed{
\Delta^{(4)}=\widehat{\Delta^{(4)}}
}
\ee
despite they appear in the denominators of very different functions,
$Ker_{[2,2,2,1^{r-6}]}$ and $\widehat{\Ker}_{[3,1^{r-3}]}$.
It is interesting, if this property (transposition independence) persists for
all higher $\Delta^{(m)}$.
}

\item{
We emphasize  that all these strangely looking combinations of the Schur functions
in $\Delta$'s are factorized at the topological locus, which explains factorization of
the associated Macdonald quantities.
The Macdonald factorizations begin to attract interest in various contexts,
see, for example, \cite{DMMSS,Konofact,CDL},
and our observations suggest that these studies can have relation to
formulas for the Kerov functions (which themselves are not factorized).
Factorizations are also crucial for knot theory applications,
and further extension of, say, the Rosso-Jones formula, which
defines the Macdonald based torus knot/link hyperpolynomials,
to the Kerov functions depends on understanding of
what substitutes the factorization in the Kerov case.
}

\end{itemize}

\section{Conclusion}

To conclude, we presented a list of properties of the Kerov functions.
Our goal was to illustrate that theory of these functions is very rich,
and they rather have {\it more} interesting properties that the Schur and Macdonald
functions, in contrary to the common belief.
Lost in the Kerov case are very special features like vanishing the multiplication structure
constants for diagrams which are not in the product of representations
and similar rigid (precise) links with representation theory,
as well as a Sugawara-like construction of the generalized cut-and-join operators
through the $U(1)$-like Cherednik-Dunkl operators.
Though important, these properties are not truly crucial for many applications,
especially because the other ones, which are really needed for practical calculations,
like the Cauchy  summation formula (\ref{Cauchy}) for the skew Kerov functions
and even the transposition rule (\ref{Ptransp}) nicely persist.
Instead of the lost properties,
an entire new world of additional time-variables $g_k$ emerges.
Moreover, a role appears for non-trivial operations over the time-variable
like inversion and multiplication, and these unexpected
operations do act on the Kerov functions,
quite non-trivially, but nicely and explicitly.

We hope that our modest review will attract an attention to this interesting field.
Particular applications were mentioned in the introduction, they will be
addressed elsewhere.
A special role among them is played by knot theory applications,
also because this theory provides one of the most promising approaches
to quantum programming \cite{Kauf,Mel}. 

\section*{Acknowledgements}

Our original interest to Kerov functions was largely inspired by Anton Zabrodin
many years ago.
We are indebted to Anton Khoroshkin for fresh comments on existing folklore
about the role and (in)significance of dominance rule
and other peculiarities of Macdonald theory.
We highly appreciate our constant discussions on related subjects with Hidetoshi Awata,
Hiroaki Kanno, Andrey Morozov, Alexei Sleptsov and Yegor Zenkevich.

This work was supported by Russian Science Foundation grant No 18-71-10073.

\section*{Appendix. Examples
\label{exa}}

Kerov functions $\widehat{\Ker}$ and $\Ker$ begin to deviate from each other only
at level $|R|=6$, which is beyond this example section, thus we do not distinguish
between them here.
At various smaller levels, we illustrate other important phenomena.

\subsection{Level 1}

Like $\Ker_\emptyset=1$ at level $|R|=0$, the polynomial at the first level is also universal,
but already its norm is $g$-dependent:
\be
\SchurP_{[1]}^{(g)} = p_1, \ \ \ ||\SchurP_{[1]}^{(g)}||^2 = g_1, \ \ \ \ \
\Ker_{[1]}\{p\} = p_1 \ \stackrel{(\ref{Ptransp})}{=}\
 ||\Ker_{[1]}^{(g)}||^2\cdot \Ker_{[1]}\left\{-\frac{p_k}{g_k}\right\}
\ee
In what follows, we omit the index $(g)$ to simplify the formulas.

\subsection{Level 2}

At level 2, there are just two functions
\be
\SchurP^{}_{[1,1]} = \Schur_{[1,1]}  = \frac{-p_2+p_1^2}{2}
\nn \\
\SchurP^{}_{[2]} = \Schur_{[2]}  + K \cdot\Schur_{[1,1]} =
\frac{(1-K)\,p_2 + (1+K)\,p_1^2}{2} = \frac{g_1^2p_2+g_2p_1^2}{g_2+g_1^2 }
= \frac{2g_2g_1^2}{g_2+g_1^2}\cdot \Schur_{[2]}\left\{-\frac{p_k}{g_k}\right\}
\ee
with the norms $||\Ker_{[1,1]}||^2={\frac{g_2+g_1^2}{2}}$ and
$||\Ker_{[2]}||^2={\frac{2g_2g_1^2}{g_2+g_1^2}}$.
\ Conditions (\ref{Ptransp}) imply that
\be
\SchurP_{[2]}\{p\} = ||\SchurP_{[2]}||^2\cdot\SchurP_{[1,1]}^\vee\{-p^\vee\}
\ \ \Longleftrightarrow \ \
\frac{g_1^2\,p_2 + g_2p_1^2}{g_2+g_1^2} = \frac{2g_1^2g_2}{g_2+g_1^2}\cdot\frac{p_2^\vee+{p_1^\vee}}{2}
\nn \\
 \SchurP_{[1,1]}\{p\} = ||\SchurP_{[1,1]}||^2\cdot\SchurP_{[2]}^\vee\{-p^\vee\} \ \ \Longleftrightarrow \ \
 \frac{-p_2+p_1^2}{2}
 =\frac{-{g_1^\vee}^2 p_2^\vee + g_2^\vee {p_1^\vee}}{2}\cdot\frac{g_1^2+g_2^2}{{g_1^\vee}^2 +g_2^\vee}
\ee
which can be easily solved:
\be
p_2^\vee = \frac{p_2}{g_2} , \ \ \ \ \ \
{p_1^\vee} = \frac{p_1}{g_1}, \ \ \ \ \ g_1^\vee = \frac{1}{g_1}, \ \ \ \ g_2^\vee = \frac{1}{g_2}
\ee
The product (\ref{Kerprod})
\be
\SchurP_{[1]}^2 = \SchurP_{[2]} + \frac{2g_1^2}{g_2+g_1^2}\cdot\SchurP_{[1,1]} \ \ \ \ \
\Longrightarrow \ \ \ \ \ \
N^{[2]}_{[1],[1]}(g)=1, \ \ \ \ N^{[1,1]}_{[1],[1]}(g) = \frac{2g_1^2}{g_2+g_1^2} =
\frac{S_{[1]}^2}{S_{[2]}}
\ee
is related to the shape of the skew Kerov functions:
\be
\SchurP_{[2]/[1]} = \frac{2g_2}{g_2+g_1^2}\cdot\Ker_{[1]} =
N^{[2]^\vee}_{[1],[1]}(g^{-1})\cdot\Ker_{[1]},
\ \ \ \ \ \
\SchurP_{[11]/[1]} = \Ker_{[1]} = N^{[1,1]^\vee}_{[1],[1]}\cdot\Ker_{[1]}
 \ee

\subsection{Level 3}

This is the first level where different $g$-dependent denominators emerge in the formulas,
but they are still related by the $g$-inversion:
\be
\Ker_{[3]} = \frac{2g_2g_1^3\,p_3+3g_3g_1^2\,p_2p_1 + g_3g_2\,p_1^3}
{2g_2g_1^2+3g_3g_1^2+g_3g_2}
= \frac{{\rm pol}(p,g)}{\Delta_3^\vee}
= \frac{\Schur_{[3]}\left\{\frac{p_k}{g_k}\right\}}{\Schur_{[3]}\left\{\frac{1}{g_k}\right\}}
\nn \\
\Ker_{[2,1]} =  \frac{-g_1(g_2+g_1^2)p_3+(g_1^3-g_3)p_2p_1+(g_3+g_2g_1)p_1^3}
{2g_3+3g_2g_1+g_1^3}
= \frac{{\rm pol}(p,g)}{\Delta_3}
\nn \\
\Ker_{[1,1,1]} = \Schur_{[1,1,1]} = \frac{p_3}{3} - \frac{p_2p_1}{2} + \frac{p_1^3}{6}
\ee
where the denominators are related by the inversion of $g$:
$\Delta_3^\vee\{g\}:=2g_2g_1^2+3g_3g_1^2+g_3g_2 = {g_3g_2g_1^3}\cdot \Delta_3\{g^{-1}\}$
with $\Delta_3\{g\} = 2g_3+3g_2g_1+g_1^3 = 6\,\Schur_{[3]}\{g_1,g_2,g_3\}$.

\bigskip

The products and skew Kerov functions are
\be
\begin{array}{c}
\Ker_{[2]}\cdot \Ker_{[1]} = \Ker_{[3]} + \\ \\ +\frac{2g_2g_1^2(2g_3+3g_2g_1+g_1^3)}{(g_2+g_1^2)(g_3g_2+3g_3g_1^2+2g_2g_1^3)}
\cdot\Ker_{[2,1]}
\\ \\ \\
 \!\!\!\!\!\!\!\!\!
 \Ker_{[1,1]}\cdot \Ker_{[1]} = \Ker_{[2,1]} + \frac{3g_1(g_2+g_1^2)}{2g_3+3g_2g_1+g_1^3}\cdot\Ker_{[1,1,1]}
\\ \\ \\
\end{array}
\!\!\!\!\!\!
\Longrightarrow \
\left\{
\begin{array}{c}
N^{[3]}_{[2],[1]}(g)=1
\\ \\
N^{[2,1]}_{[2],[1]}(g) = \frac{2g_2g_1^2(2g_3+3g_2g_1+g_1^3)}{(g_2+g_1^2)(g_3g_2+3g_3g_1^2+2g_2g_1^3)}  \\ \\ \\
N^{[2,1]}_{[1,1],[1]}(g)=1
\\ \\
N^{[1,1,1]}_{[1,1],[1]}(g) = \frac{3g_1(g_2+g_1^2)}{2g_3+3g_2g_1+g_1^3}   =
\frac{\Schur_{[2]}\{g\}\cdot\Schur_{[1]}\{g\}}{\Schur_{[3]}\{g\}}
\!\!\!\!\!\!\!\!\!
\end{array}
\right.
\ee
\be
\Ker_{[3]/[2]} = N^{[1,1,1]}_{[1,1],[1]}(g^{-1})\cdot \Ker_{[1]}, \ \ \ \ \ \ \ \ \
\Ker_{[3]/[1]} = N^{[1,1,1]}_{[1,1],[1]}(g^{-1})\cdot \Ker_{[2]}
\nn \\
\Ker_{[2,1]/[2]} = N^{[2,1]}_{[1,1],[1]}(g^{-1})\cdot \Ker_{[1]}, \ \ \ \ \ \ \ \ \
\Ker_{[2,1]/[1,1]} = N^{[2,1]}_{[2],[1]}(g^{-1})\cdot \Ker_{[1]} = \Ker_{[1]},
\nn \\
\Ker_{[2,1]/[1]} = N^{[2,1]}_{[1,1],[1]}(g^{-1})\cdot \Ker_{[2]} + N^{[2,1]}_{[2],[1]}(g^{-1})\cdot \Ker_{[1,1]}
= \Ker_{[2]} + N^{[2,1]}_{[2],[1]}(g^{-1})\cdot \Ker_{[1,1]}
\nn \\
\Ker_{[1,1,1]/[1,1]} = N^{[3]}_{[2],[1]}(g^{-1})\cdot \Ker_{[1]} = \Ker_{[1]}, \ \ \ \ \ \ \ \ \
\Ker_{[1,1,1]/[1]} = N^{[3]}_{[2],[1]}(g^{-1})\cdot \Ker_{[1,1]} = \Ker_{[2]}
\ee
Note also that
\be
\Delta_3\cdot \Ker_{[2,1]} = -g_1p_3\cdot \Delta_{2}  +
{g_1p_1}\cdot {\Delta_2} \cdot \Ker_{[2]} + 2g_3p_1\cdot \Ker_{[1,1]}
\label{Ker21}
\ee

\subsection{Level 4}

Expressions for various denominators $\Delta_4\{g\}$ will be provided
in the next subsection. Other quantities are

\be
\Ker_{[4]} = \frac{6g_3g_2^2g_1^4\cdot p_4 + 8g_4g_2^2g_1^3\cdot p_3p_1
+3g_4g_3g_1^4\cdot p_2^2+  6g_4g_3g_2g_1^2\cdot p_2p_1 +
g_4g_3g_2^2\cdot p_1^4}{\Delta_{4}^\vee}
= \frac{\Schur_{[4]}\left\{\frac{p_k}{g_k}\right\}}{\Schur_{[4]}\left\{\frac{1}{g_k}\right\}}
\nn \\ \nn \\
\Ker_{[3,1]} =   \frac{{\rm pol}(p,g)}{2\Delta'^\vee_{4} },   \ \ \ \ \
\Ker_{[2,2]} = \frac{{\rm pol}(p,g)}{2\Delta'_{4} }, \ \ \ \ \
\Ker_{[1,1,2]} = \frac{{\rm pol}(p,g)}{2\Delta_4 }
\nn \\ \nn \\
\Ker_{[1,1,1,1]} = \Schur_{[1,1,1,1]}\{p_k\}
\ee

The simplest of the three polynomials in the numerators is in
{\footnotesize
$$
\!\!\!\!\!\!\!\!\!\!\!\!\!\!\!\!
\Ker_{[1,1,2]} =
\frac{g_1\overbrace{(2g_3+3g_2g_1+g_1^3)}^{\Delta_{3}}\cdot( 2p_4 -p_2^2) + 2(2g_4+g_2^2-2g_2g_1^2-g_1^4)\cdot p_3p_1
- (6g_4+4g_3g_1+3g_2^2-g_1^4)\cdot p_2p_1^2+ (2g_4+2g_3g_1+g_2^2+g_2g_1^2)\cdot p_1^4}{2\Delta_{4}}
$$
}

\noindent
but they are hardly useful, if presented in this form.
Remarkably, this complicated expression is nothing but a very simple
(\ref{Keradj}), which is a direct generalization of  (\ref{Ker21}).

\subsection{Examples of the emerging structure}

We can now list the emerging denominators
and observe that they are actually the Schur functions with $g_k$ playing the role
of the time variables:

\be
\begin{array}{ll}
\Delta_2 = &g_2+g_1^2 = \boxed{2\,\Schur_{[2]}\{g_1,g_2\}}
\\   \\   \\
\Delta_3  = &2g_3+3g_2g_1+g_1^3 =\boxed{ 6\,\Schur_{[3]}\{g_1,g_2,g_3\}},
  \\
\Delta_3^\vee=&g_3g_2+3g_3g_1^2+2g_2g_1^3 =
 \boxed{6g_3g_2g_1^3\cdot \Schur_{[3]}(g_1^{-1},g_2^{-1},g_3^{-1})}
=   \\   \\
&= 6{\Big(\Schur_{[3]}\cdot\Schur_{[2]}+\Schur_{[3]}\cdot\Schur_{[1]}^2
-\Schur_{[2]}^2\cdot\Schur_{[1]}\Big)\{g\}} \sim
  \\
& \ \ \ \ \
\sim \Big([32]+[311]-[221]\Big)
= \det\left(\begin{array}{ccc}
\Schur_{[3]} & 0 & \Schur_{[1]} \\
\Schur_{[3]} & \Schur_{[2]} & 0 \\
\Schur_{[2]}  & \Schur_{[1]} & 1
\end{array} \right)^{\phantom{^5}}
 \\   \\  \\
\Delta_4 = &6g_4+8g_3g_1+3g_2^2 +6g_2g_1^2+g_1^4   = \boxed{24\,\Schur_{[4]}\{g_k\}}
  \\   \\
\Delta_4^\vee= &g_4g_3g_2^2+6g_4g_3g_2g_1^2 + 8g_4g_2g_1^2+3g_4g_3g_1^4+6g_3g_2^2g_1^4
\sim \boxed{\Schur_{[4]}\{g^{-1}_k\}}
  \\   \\
\Delta'_{4} =&2\Big(2g_4g_3+g_3g_2^2+6g_4g_2g_1+ 3g_2^3g_1+ 2g_3g_2g_1^2+ 4g_4g_1^3
+ 2g_2^2g_1^3+3g_3g_1^4+g_2g_1^5\Big)
=   \\
&= 48\Big(\Schur_{[4]}\cdot\Schur_{[3]} + \Schur_{[4]}\cdot\Schur_{[2]}\cdot\Schur_{[1]}
- \Schur_{[3]}^2\cdot\Schur_{[1]}\Big)\{g\}
\sim   \\
& \ \ \ \ \
\sim\Big([43]+[421]-[331]\Big)
= \det\left(\begin{array}{ccc}
\Schur_{[4]} & 0 & \Schur_{[1]} \\
\Schur_{[4]} & \Schur_{[3]} & 0 \\
\Schur_{[3]}  & \Schur_{[2]} & 1
\end{array} \right)^{\phantom{^5}}
  \\  \\
\Delta'^\vee_{4} = &\Delta'_{4}\{g_k^{-1}\}
\sim\Big(
\underline{2[4322]+[431111]}+2[42221]-4[422111]+[4211111] -
 \\
&\ \ \ \ \ \ \ \ \ \
- 2[33221]-[3311111]-[32222]+4[322211]-[222221]\Big)

\end{array}
\label{Deltalist}
\ee
Explicitly written in
(\ref{Delta4}) is the lifting to arbitrary $r$
of just the two underlined terms from the last formula.
Note that the degeneracy is lifted for $r>4$, and they become three independent structures.
The same is going to happen to the other items, thus, in general, it is a 19-term expression.

This form of $\Delta_r$ can explain why the Macdonald choice of $g_k$ is distinguished:
it is an exact counterpart of the topological locus, which converts the Schur functions
into the quantum dimensions $g_k=\frac{\{A^k\}}{\{t^k\}}$.
However, the factorization of other denominators is far from obvious.
What if we further change the Schur functions of $g_k$ for the Macdonald ones and then to the Kerov functions?

Since this Appendix is devoted to honest examples,
which illustrate and {\it underlie} general theory rather than follow from it,
we do not interpret the denominators $\Delta$ in the terms of sec.\ref{gvars},
instead we present them as they are.
Actually, in general terms (i.e. for all $r$),
\be
\begin{array}{cc}
\text{second\ diagram}  &
\Delta_r = \Delta_{[2,1^{r-2}]} = \Delta^{(2)}_r\{g_k\} \\ \\
\text{third\ diagram}  &
\Delta'_r = \Delta_{[2,2,1^{r-2}]} = \Delta^{(3)}_r\{g_k\} \\ \\
\text{fourth diagram}  &
\Delta''_r = \Delta_{[2,2,2,1^{r-6}]}=\widehat{\Delta}_{[3,1^{r-3}]} =
\Delta^{(4)}_r\{g_k\}
\\ \\ \ldots \\ \\
\text{third-from-the end diagram} &
\Delta''^\vee_r = \Delta_{[r-2,2]} = \Delta''_r\{g_k^{-1}\}\\ \\
\text{penultimate diagram} & \Delta'^\vee_r = \Delta_{[r-1,1]} = \Delta'_r\{g_k^{-1}\}\\ \\
\text{last diagram} & \Delta^\vee_r = \Delta_{[r]} = \Delta_r\{g_k^{-1}\}
\end{array}
\ee
Note that the lexicographically-last diagram is related by the $g$-inversion
to the {\it second} diagram, not to the first one, and so on:
\ there is a {\it shift-by-one}
in these formulas as compared to a more naive expectation from (\ref{Ptransp}).
In result, while the "true" fourth diagram appears only at level $6$,
its counterpart is the third diagram from the end, and it is non-trivial
already at level $4$, which allows one to learn something about $\Delta^{(4)}$
from the example at this level.
The structure which is revealed in this way is indeed true in general, see (\ref{Delta4})
in sec.\ref{gvars}.

\end{document}